\documentclass[letter]{aa}

\usepackage{graphicx}
\usepackage{txfonts}
\usepackage{amsmath}
\usepackage[T1]{fontenc}
\usepackage{microtype}
\usepackage{tikz}
\usetikzlibrary{arrows.meta,positioning,calc}

\newcommand{\dd}{\mathrm d}
\newcommand{\vect}[1]{\boldsymbol{#1}}
\newcommand{\mat}[1]{\widehat{\mathsf{#1}}}

\begin{document}
\raggedbottom

\title{Origin of the angular momentum of filaments in the local Universe}

\author{V.G.Gurzadyan\inst{1,2}, N.N.Fimin\inst{3}, V.M.Chechetkin\inst{3} }

              \institute{Center for Cosmology and Astrophysics, Alikhanian National
Laboratory and Yerevan State University, 0025 Yerevan, Armenia \and
SIA, Sapienza University of Rome, 00191 Rome, Italy \and Keldysh Institute of Applied Mathematics of RAS, 125047 Moscow, Russia}

 \offprints{V.G. Gurzadyan, \email{gurzadyan@yerphi.am}}
 \date{Submitted: XXX; Accepted: XXX}

\abstract
{Velocity gradients compatible with rotation have been reported across certain nearby cosmic filaments.  We show that rotational momentum might be generated in filaments in the local Universe in the presence of a cosmological constant (local dark energy) based on the theorem stating the general function for the equivalence of the gravitational field of a sphere and that of a point mass.
We investigated how the cosmological constant modifies the finite-time acquisition and retention of angular momentum in an open filamentary core. Although its isotropic force exerts no direct axial torque, rotational motion can be seeded by tidal interactions with surrounding matter and incoming streams. Early tidal torques establish an imbalance in axial angular momentum, while the subsequent orbit crossing amplifies the resulting rotation, and ultimately, the cosmological constant is able to modify tidal-mode growth and finite-time angular momentum delivery.
} 


   \authorrunning{V.G. Gurzadyan, N.N.Fimin, V.M.Chechetkin}
   \titlerunning{Origin of the angular momentum of filaments in the local Universe}
   \maketitle
%\maketitle

\section{Introduction}

The claimed detection of a 15 Mpc rotating filament at $z=0.032$ \citep{Tudorache2025}, along with earlier possibly similar evidence (e.g., \citep{Wang2021,Tang2025,Wang2025}), add remarkable features to the matter distribution in the local Universe. 
Although it remains unclear whether filaments with rotational momentum are common or unique, and although far more extensive observational surveys are required to characterize their detailed properties, the existing observational evidence motivates us to explore possible mechanisms for their formation and evolution.

The Hubble tension and other observational evidence \citep{R1,R2,R3,LR,Mon,DESI1,DES,Dai,Cas,Ban1,Jia,Cap} are considered as possible genuine differences in the descriptions of the local (late) and the early Universe. 
The rotational motions of cosmic filaments provide a complementary probe of late-time structure formation, and in previous studies, the tidal-torque acquisition \citep{White1984}, spin alignments \citep{Codis2015}, and multi-stream vorticity \citep{Pueblas2009} of gravitating systems were considered. We examine how the cosmological constant can modify these processes, namely, the angular momentum transfer to a filament core.

The differences between the processes occurring in the local and early Universe, including the Hubble tension, which is interpreted as a consequence of two flows (a local and a global flow), were considered in \citep{GS6,GS7,GS8}. These authors derived absolute maximum and minimum limits for the local Hubble parameter that fully agreed with observational data. This approach was based on a theorem \citep{G1} (for a pedagogical discussion, see \citep{Car}) stating the general force law satisfying the identity between the gravitational field of a sphere and of a point mass in the form
\begin{equation}
F=-{\frac{G Mm}{r^{2}}}+{\frac{\Lambda c^{2}mr}{3}}.
\end{equation}
Within weak-field General Relativity and in nonrelativistic cosmological models at local scales \citep{MM,Z81} $\Lambda$ in Eq.(1) corresponds to the cosmological constant. It is important to note that unlike the Newtonian $1/r^2$ law, the law given by Eq. (1) does not lead to a force-free field inside a shell. There are observational indications that the disks of spiral galaxies are affected by the spherical halos of their host galaxies \citep{Kr}. \citep{PRR} considered a tabletop-scale experimental test of this law.

The properties of groups and clusters of galaxies in the local Universe were shown to fit Eq.(1) ( \citep{G2,GS2,GS3,GS4,GS5}; see also \citep{RG} regarding the vicinity of the Local Group).
The properties of the voids within the same approach with the cosmological constant were studied in \citep{GFC1,GFC2,GFC3,GFC4,GFC5}. 

Below, we consider the possibility of filament formation with nonzero angular momentum in the local Universe using the approach based on Eq. (1), which incorporates the cosmological constant.
We show that the $\Lambda$ term modifies tidal trajectories and stream arrival times, and the boundary flux, matter torque, and evolving moment of inertia determine the sign and size of the resulting spin-up.
 
\section{The momentum seed and its tidal amplification}

We considered a fixed cylindrical surface surrounding a short, nearly straight filament segment. Two cold, collisionless streams were followed from an upstream surface to this cylinder. The streams did not interact at the boundary; each crossed it with its own density and velocity. The cylinder was able to gain angular momentum only through the flux carried across the boundary or through the gravitational torque exerted on matter already inside,
\begin{equation*}
 J_{\rm in}\ \longrightarrow\
 (\Delta u_n,\Delta u_\phi)\ \longrightarrow\
 \dot L_z\ \longrightarrow\ \Omega=\frac{L_z}{I}.
\end{equation*}
The first arrow is the tidal change before arrival. The second arrow is the crossing of the cylindrical boundary. The last arrow is the conversion from angular momentum input to angular acceleration. 

On a fixed upstream surface $\Sigma_{\rm in}$, let $\rho_s$ be the density of an incoming cold stream $s$, $u_{n,s}<0$ its inward normal velocity, $u_{\phi,s}$ its velocity around the filament axis, and $r_\perp$ its distance from that axis. The signed axial angular momentum input is
\begin{equation}
 J_{\rm in}(t)=\sum_s\int_{\Sigma_{\rm in},\,u_{n,s}<0}
 r_\perp\rho_su_{\phi,s}|u_{n,s}|\,\dd S .
 \label{eq:Jin}
\end{equation}
The factor $r_\perp u_{\phi,s}$ is the specific axial angular momentum, while $\rho_s|u_{n,s}|\dd S$ is the mass-crossing rate. Large positive and negative terms can nearly cancel. Their small remainder gives the upstream rotational orientation. It does not by itself fix the amplitude or even guarantee the sign later deposited at the core, because the streams follow different paths before they arrive. The incoming remainder fixes only the upstream sign; the boundary flux fixes what enters the core.

We therefore selected two small streams, denoted by $b$ and $c$, and followed the same material as it moved toward the core. At the boundary, we specified their densities and all components of their velocities. It is essential to follow a small stream rather than a single trajectory because the density changes as neighboring trajectories converge or diverge.

Let $\Phi_m$ denote the matter potential. When the cosmological term is included in Eq.~(1), the local potential and the equation of motion can be written as
\begin{equation}
 \begin{aligned}
 \nabla^2\Phi_m&=4\pi G\rho_m,\qquad
 \Phi=\Phi_m-\frac12\omega_\Lambda^2 r^2,\\
 \ddot{\vect r}&=-\nabla\Phi_m+\omega_\Lambda^2\vect r,
 \qquad \omega_\Lambda^2=\frac{c^2\Lambda}{3}.
 \end{aligned}
 \label{eq:motion}
\end{equation}
 The additional acceleration is isotropic. It does not select a rotation direction, but changes the relative motion of nearby trajectories.
 For trajectories $\vect r(t)$ and $\vect r(t)+\vect\xi(t)$ in the same smooth potential, subtraction gives exactly
\begin{equation}
\begin{aligned}
\ddot{\vect\xi}={}&-\left[\nabla\Phi_m(\vect r+\vect\xi,t)-\nabla\Phi_m(\vect r,t)\right]\\
&+\omega_\Lambda^2\vect\xi .
\end{aligned}
\tag{3a}\label{add:separation}
\end{equation}
When the separation is small relative to the scale of force variation,
\begin{equation}
\begin{aligned}
\nabla\Phi_m(\vect r+\vect\xi,t)-\nabla\Phi_m(\vect r,t)
 &=\mat H\vect\xi+O(|\vect\xi|^2),\\
H_{ij}&=\partial_i\partial_j\Phi_m .
\end{aligned}
\tag{3b}\label{add:linearisation}
\end{equation}
Here, $\mat H$ is evaluated along $\vect r(t)$, and $\mat I$ is the identity matrix. Its eigenvalues obey $\sum_i\mu_i=4\pi G\rho_m$. 

We defined the matter tidal matrix by $H_{ij}=\partial_i\partial_j\Phi_m$. For a small separation $\vect\xi$ between neighboring trajectories, linearization gives
\begin{equation}
 \ddot{\vect\xi}=-\bigl(\mat H-\omega_\Lambda^2\mat I\bigr)\vect\xi .
 \label{eq:Jacobi}
\end{equation}
Along a principal direction, with the matter curvature $\mu_i$, the Jacobi equation becomes
\begin{equation}
 \ddot\xi_i+(\mu_i-\omega_\Lambda^2)\xi_i=0,
 \qquad
 \sigma_i=\sqrt{\omega_\Lambda^2-\mu_i}
 \quad (\mu_i<\omega_\Lambda^2).
 \label{eq:mode}
\end{equation}
The meaning is simple. If $\mu_i>\omega_\Lambda^2$, the motion in that direction is oscillatory. If $\mu_i<\omega_\Lambda^2$, it contains a growing part. The cosmological term lowers the restoring curvature by the same amount in every direction. It therefore increases a rate that is already positive, and more importantly, it can create a growing direction when $0<\mu_i<\omega_\Lambda^2$. In this interval the matter tide alone is restoring, whereas the combined equation is growing.

The decoupled model interpretation of Eq.~(\ref{eq:mode}) assumes that the curvatures and their principal directions are approximately constant over $T=t-t_0$. Instantaneous eigenvalues alone do not decouple a time-dependent rotating eigenbasis. With $\xi_{i0}=\xi_i(t_0)$ and $v_{i0}=\dot\xi_i(t_0)$, the growing-case solution is
\begin{equation}
\begin{aligned}
\xi_i(t)&=\xi_{i0}\cosh(\sigma_iT)
 +\frac{v_{i0}}{\sigma_i}\sinh(\sigma_iT),\\
\dot\xi_i(t)&=\sigma_i\xi_{i0}\sinh(\sigma_iT)
 +v_{i0}\cosh(\sigma_iT).
\end{aligned}
\tag{5a}\label{add:fullsolution}
\end{equation}
A growing term is excited only if $\xi_{i0}+v_{i0}/\sigma_i\ne0$. For the restoring case, we write $\nu_i=\sqrt{\mu_i-\omega_\Lambda^2}$. Then
\begin{equation}
\begin{aligned}
\xi_i(t)&=\xi_{i0}\cos(\nu_iT)
 +\frac{v_{i0}}{\nu_i}\sin(\nu_iT),\\
\dot\xi_i(t)&=-\nu_i\xi_{i0}\sin(\nu_iT)
 +v_{i0}\cos(\nu_iT).
\end{aligned}
\tag{5b}\label{add:oscsolution}
\end{equation}
At the threshold, $\xi_i=\xi_{i0}+v_{i0}T$ and $\dot\xi_i=v_{i0}$. These finite-time solutions, with the same initial data, provide the appropriate comparison across a tidal threshold. The relation $\Delta\vect u=\dot{\vect\xi}$ applies to the nearby trajectories only while the linear approximation remains valid.

The normal and azimuthal directions at the cylinder need not be principal tidal directions. It is sufficient that the growing motions have nonzero projections on them. In the simplest case, when the two velocity differences are dominated by growing projections with rates $\sigma_n$ and $\sigma_\phi$,
\begin{equation}
 \Delta u_n\Delta u_\phi\simeq
 (\Delta u_n\Delta u_\phi)_0\,e^{\gamma_\Lambda\tau},
 \qquad
 \gamma_\Lambda\simeq\sigma_n+\sigma_\phi .
 \label{eq:velocitygrowth}
\end{equation}
Here, $\tau$ is the time spent by the selected material in the growing part of the tidal field. The formula is not needed as a universal identity: its purpose is to show where the exponential dependence can enter. The actual values of $\Delta u_n$ and $\Delta u_\phi$ are the velocities carried by the streams when they reach the cylinder. The cosmological term changes the gain, not the rotational sign.

The physical arrival time $t$ need not coincide with a common amplification time $\tau$: different cohorts may have different residence times, density weights, and cancellations. Thus, Eq.~(\ref{eq:velocitygrowth}) is a statement about a local mode, not yet a law for the full signed core input.

The boundary values require a Lagrangian map for each cold stream, $\vect r_s(\vect q,t)$, where $\vect q$ is its initial physical position. For a local pair, the pre-images must satisfy $\vect r_b(\vect q_b,t)=\vect r_c(\vect q_c,t)\in\Sigma_c$. The growth of a separation by itself establishes neither this meeting nor the arrival times. 

On each regular branch, mass conservation gives
\begin{equation}
\partial_t\rho_s+\nabla\!\cdot(\rho_s\vect u_s)=0,
\tag{6a}\label{add:continuity}
\end{equation}
\begin{equation}
\rho_s(\vect r,t)=\frac{\rho_{s0}(\vect q_s)}{|J_s(\vect q_s,t)|},
\quad J_s=\det\!\left(\frac{\partial\vect r_s}{\partial\vect q}\right).
\tag{6b}\label{add:densitymap}
\end{equation}
Here, $\vect q_s$ is the relevant pre-image of $\vect r$ and $J_s\ne0$. Distinct pre-images contribute distinct streams; the formula is not a finite point-wise density prescription at a caustic where $J_s=0$. Finite local densities require a specified velocity dispersion or an explicit averaging prescription.

\section{Angular momentum through the core boundary}

Let $V_c$ be the fixed cylindrical core and $\Sigma_c=\partial V_c$ its boundary, with outward normal $\vect n$. The exact axial angular momentum balance is
\begin{equation}
 \dot L_z=-\oint_{\Sigma_c}r_\perp\sum_s\rho_su_{\phi,s}u_{n,s}\,\dd S
 -\int_{V_c}\rho\,\partial_\phi\Phi_m\,\dd^3r .
 \label{eq:budget}
\end{equation}
The first term is the angular momentum carried through the cylinder, and the second term is the gravitational torque on matter already inside. 

Because $d\ell_z/dt=-\partial_\phi\Phi_m$, $\Lambda$ exerts no direct axial torque. In an axisymmetric matter potential, identical initial particles retained to completion contribute the same final $L_z$ for any $\Lambda$. At a fixed observation time, however, $\Lambda$ can alter which particles have entered or remain in the core; in a non-axisymmetric potential, the matter torque also acts along $\Lambda$-dependent trajectories.

At a given point of the boundary, we define $\Delta u_j=u_{j,b}-u_{j,c}$ and $\bar u_j=(\rho_bu_{j,b}+\rho_cu_{j,c})/(\rho_b+\rho_c)$. The complete pair flux decomposes exactly as
\begin{equation}
 %\begin{aligned}
\sum_{s=b,c}\rho_su_{\phi,s}u_{n,s}=(\rho_b+\rho_c)\bar u_\phi\bar u_n\,
\quad +W\Delta u_\phi\Delta u_n,\quad W=\frac{\rho_b\rho_c}{\rho_b+\rho_c}.
 %\end{aligned}
\label{eq:pairflux}
\end{equation}

For a crossed area $A(t)$ near $r_c$, the relative contribution only is
\begin{equation}
\dot L_{z,\mathrm{rel}}^{bc}\simeq-r_cA(t)W(t)\Delta u_n(t)\Delta u_\phi(t).
\label{eq:pairinput}
\end{equation}
For the complete pair, we must also integrate the mean term in Eq.~(\ref{eq:pairflux}); other streams and $-\int_{V_c}\rho\,\partial_\phi\Phi_m\,\dd^3r$ enter the full core budget. Mass conservation gives $\rho_sA_s|u_{n,s}|=\dot M_s$ for a narrow steady crossing stream, and hence, its signed inward contribution is $\dot M_s\ell_s$, $\ell_s=r_cu_{\phi,s}$.

\section{Spin-up of the filament core}

A constructive full-flux example is given by two cold sheets entering a small cylinder patch $x=R$: $x_s=X-U_st$, $y_s=Y\cosh(\sigma t)+\epsilon_s(v_*/\sigma)e^{\sigma t}$ with $\epsilon_b=-\epsilon_c=1$, $0<\sigma<\omega_\Lambda$, and a static linear matter tide $H_{xx}=\omega_\Lambda^2$, $H_{yy}=\omega_\Lambda^2-\sigma^2$. For initial labels symmetric in $Y$, their densities and crossed areas are $\rho_s=\rho_0/\cosh(\sigma t)$ and $A_s=A_0\cosh(\sigma t)$, so $\rho_sA_sU_s=\rho_0A_0U_s$ is fixed. Integration of the complete $\ell_z=x\dot y-y\dot x$ yields
\begin{equation}
\begin{aligned}
\dot L_{z,\mathrm{in}}={}&\rho_0A_0v_*e^{\sigma t}\bigl[U_b(R+U_b/\sigma)\\[-2pt]
&-U_c(R+U_c/\sigma)\bigr].
\end{aligned}
\label{eq:full_example}
\end{equation}

We completed the three-dimensional map with $z=Z$ and $H_{zz}=\omega_\Lambda^2$: then $J_s=\cosh(\sigma t)$, $\rho_sA_s=\rho_0A_0$, and $4\pi G\rho_m=3\omega_\Lambda^2-\sigma^2>0$ for $0<\sigma<\omega_\Lambda$. The plane $x=R$ approximates a cylinder patch only while $|y|/R\ll1$ and normal-flux curvature corrections are negligible. Arrivals at $t$ started at $X=R+U_st$, giving an exponential complete signed boundary input, not an assumed history. The exponential net core $L_z$ remains conditional on its internal matter torque.

Spin-up means an increase in the angular velocity, not only an increase in angular momentum. Let
$$
 I(t)=\int_{V_c}\rho r_\perp^2\,\dd^3r,
 \qquad
 \Omega(t)=\frac{L_z(t)}{I(t)},
 \qquad
 \dot\Omega=\frac{\dot L_z}{I}-\Omega\frac{\dot I}{I}.
 \label{eq:Omega}
$$

The first term is the signed contribution from the full boundary flux and internal matter torque, and it need not accelerate the rotation. Accretion at large radius may increase $I$ and reduce $|\Omega|$ even if $|L_z|$ grows. For $L_z>0$,

$$
 \dot\Omega>0
 \quad\Longleftrightarrow\quad
 \frac{\dot L_z}{L_z}>\frac{\dot I}{I}.
 \label{eq:spincondition}
$$
This inequality is the direct condition for rotational acceleration.

This equation assumes $\Omega>0$, not merely a fixed sign. For $I>0$ and either nonzero sign of $L_z$, the growth of the rotation speed obeys
\begin{equation}
\frac{1}{|\Omega|}\frac{\dd|\Omega|}{\dd t}
=\frac{\dot L_z}{L_z}-\frac{\dot I}{I}>0 .
%\tag{11a}\label{add:spinsign}
\end{equation}
At $L_z=0$, we used the non-singular differential identity in Eq.~(\ref{eq:Omega}), not this logarithmic form. The quantity $L_z/I$ is an inertia-weighted effective angular velocity; it is not evidence of rigid or spatially coherent rotation.

For the same fixed control volume, continuity also gives
\begin{equation}
\begin{aligned}
\dot I={}-\oint_{\Sigma_c}r_\perp^2
       \sum_s\rho_su_{n,s}\,\dd S
 +2\int_{V_c}r_\perp\sum_s\rho_su_{r,s}\,\dd^3r ,
\end{aligned}
%\tag{11b}\label{add:inertiabudget}
\end{equation}
where $u_{r,s}$ is the cylindrical radial velocity. Thus, influx, outflux, and radial redistribution must also be evolved when converting an angular momentum history into spin-up. 
Equation~(\ref{eq:pairinput}) contains only the relative-motion contribution. The total input also depends on the mean stream transport, other streams, and any matter torque. Neither its sign nor its time dependence follows from that product alone.

Let $\Delta t=t-t_c$ and neglect other fluxes and the volume torque during the short onset. For a stream band whose straight leading edge of length $\ell$ sweeps across the cylindrical surface at speed $v_e$, $A(t)\simeq\ell v_e\Delta t$. If the complete signed local flux density $j_c=-r_c\sum_{s=b,c}\rho_su_{\phi,s}u_{n,s}>0$ is approximately constant, then
\begin{equation}
\dot L_z\simeq K\Delta t,\qquad L_z-L_z(t_c)\simeq\frac12K\Delta t^2,\quad K=A_1j_c.
\label{eq:onset}
\end{equation}
A finite initial crossed area gives a linear onset. Different contact geometries may yield other powers.

Linear accumulation requires a constant net $\dot L_z$, including both the complete signed boundary flux and the volume torque. If the complete local signed flux density instead scales as $j_c(t)=j_c(t_c)e^{\gamma_\Lambda\Delta t}$ while $A\simeq A_1\Delta t$ and other budget terms are negligible, the conditional model is
\begin{equation}
\dot L_z=K\Delta t\,e^{\gamma_\Lambda\Delta t},\quad L_z-L_z(t_c)=\frac{K}{\gamma_\Lambda^2}\left[e^{\gamma_\Lambda\Delta t}(\gamma_\Lambda\Delta t-1)+1\right].
\label{eq:combined}
\end{equation}
After the area saturates, the corresponding net-rate assumption gives
\begin{equation}
\dot L_z=\dot L_c e^{\gamma_\Lambda\Delta t},\quad L_z-L_z(t_c)=\frac{\dot L_c}{\gamma_\Lambda}(e^{\gamma_\Lambda\Delta t}-1).
\label{eq:expstage}
\end{equation}
These are conditional complete-input histories, not necessary consequences of the Jacobi exponents. If the full signed rate remains constant, $L_z$ grows linearly regardless of the amplification experienced by individual trajectories.
The sign of the complete input, not only that of $\Delta u_n\Delta u_\phi$, determines whether $L_z$ increases or decreases; cancellations can suppress a growing relative contribution.

For a slowly changing $I$, the same law gives the angular velocity directly,
\begin{equation}
 \Omega(t)-\Omega(t_c)\simeq
 \frac{\dot L_c}{I_c\gamma_\Lambda}
 \left(e^{\gamma_\Lambda\Delta t}-1\right).
 \label{eq:Omegagrowth}
\end{equation}
When the exponential part dominates, $L_z$ and $\Omega$ both change on the timescale $\gamma_\Lambda^{-1}$. Before this, the opening of the crossed area controls the rise. If $I$ changes appreciably, the exact condition remains $\dot L_z/L_z>\dot I/I$.
When $I$ changes, the evolution of $L_z/I$ must be evaluated from Eq.~(\ref{eq:Omega}). A positive change in $L_z$ alone does not ensure an increase in the rotation speed.

For mirror-symmetric galaxy samples, $\Delta v_{\rm los, rot}\simeq 2\sin i\,\langle u_\phi|\cos\varphi|\rangle_{\rm gal}$, with $i$ the axis--sightline angle, and $\varphi$ the azimuth from maximum projection. The simplification $2\langle u_\phi\rangle_{\rm gal}\sin i$ assumes sampling near that azimuth. The viewing geometry, galaxy selection, dynamical state and filament identification \citep{Wang2021,Tang2025} affect the signal, which differs from matter-weighted $L_z/I$.

\section{Cumulative role of the cosmological constant} 

At fixed matter tidal curvatures $\mu_i$, in a two-growing-mode approximation, if the complete signed input obeys Eq.~(\ref{eq:expstage}) for both $\Lambda=0$ and $\Lambda>0$, then

\begin{equation}
 %\begin{aligned}
 \frac{\partial\gamma_\Lambda}{\partial\Lambda}
 \simeq\frac{c^2}{6}\left(\frac1{\sigma_n}+\frac1{\sigma_\phi}\right)>0,\,\,
 \frac{\dot L_{z,\Lambda}}{\dot L_{z,0}}
 \simeq\frac{\dot L_{c,\Lambda}}{\dot L_{c,0}}\,
 e^{(\gamma_\Lambda-\gamma_0)\Delta t}.
 %\end{aligned}
 \label{eq:lambdaeffect}
\end{equation}

The pre-factors in Eq.~(\ref{eq:lambdaeffect}) need not agree between the models. When $0<\mu_i<\omega_\Lambda^2$, a restoring mode without $\Lambda$ starts to grow with it. Nevertheless, the growing rates alone do not bound the complete signed flux. The scale $\omega_\Lambda^{-1}\simeq17.5$ Gyr makes the finite inflow duration essential, and the finite-time Jacobi solution remains regular at the tidal threshold.
 
As an illustration, we considered a uniform cylindrical matter potential $\Phi_m=\kappa^2r_\perp^2/2$ over $r_\perp\le r_0=1\,\mathrm{Mpc}$ with $\kappa=0.10\,\mathrm{Gyr}^{-1}$ and $r_c=0.30\,\mathrm{Mpc}$. Thus, $\rho_m=\kappa^2/(2\pi G)\simeq3.5\times10^{11}M_\odot\,\mathrm{Mpc}^{-3}$ and a $1$ Mpc core segment has $\sim10^{11}M_\odot$. Two test streams launch continuously at $r_0$ with $(u_{r,+},u_{r,-})=(-0.20,-0.15)\,\mathrm{Mpc\,Gyr}^{-1}$, $\ell_\pm=\pm0.03\,\mathrm{Mpc}^2\,\mathrm{Gyr}^{-1}$, and $\dot M_\pm=(5,4.5)M_\odot\,\mathrm{yr}^{-1}$. The orbital timescale is denoted by $a=t-t_{\rm inj}$. 
At equal initial conditions, $\alpha^2=\kappa^2-\omega_\Lambda^2$ gives the exact planar orbit,
\begin{equation}
%\begin{aligned}
r_s^2(a)=\left[r_0\cos(\alpha a)+\frac{u_{r,s}}{\alpha}\sin(\alpha a)\right]^2
+\left[\frac{\ell_s}{r_0\alpha}\sin(\alpha a)\right]^2.
%\end{aligned}
\label{eq:orbits}
\end{equation}
The first-crossing timescales are $(3.367,4.366)$ Gyr for $\Lambda=0$ and $(3.437,4.522)$ Gyr for $\omega_\Lambda^{-1}=17.5$ Gyr. Both streams lie in the invariant plane $z=u_z=0$, so the endcap flux vanishes. No radial recrossings occur before $T=5\,\mathrm{Gyr}$. The prescribed initially non-rotating background has fixed $I_c=M_cr_c^2/2$. The steady inflows give $L_z(T)=\sum_s\dot M_s\ell_s[T-\tau_s]_+$, equal to $1.594$ and $1.699\times10^8\,M_\odot\,\mathrm{Mpc}^2\,\mathrm{Gyr}^{-1}$ ($+6.6\%$). With $I(T)=M_cr_c^2/2+\sum_s\dot M_s\int_{\tau_s}^{T}r_s^2(a)\,\mathrm da$ for $M_c=10^{11}M_\odot$, the corresponding effective angular velocities are $31.3$ and $33.6\,\mathrm{km\,s}^{-1}\,\mathrm{Mpc}^{-1}$ ($+7.3\%$), respectively. These parameters describe predominantly radial infall, for instance, $150$--$200\,\mathrm{km\,s^{-1}}$, versus $30\,\mathrm{km\,s^{-1}}$ tangentially at $r_0$. Similar counter-stream rates illustrate incomplete cancellation. The fixed-background test-stream approximation neglects stream self-gravity and background response. The injected mass is $10$--$11\%$ of $M_c$ at $T=5\,\mathrm{Gyr}$. The quoted increases are illustrative and not precision predictions for a self-consistent filament.

Finally, let $t_s$ be the time at which the complete signed input in Eq.~(\ref{eq:budget}) ends. Then
\begin{equation}
 L_z(t\ge t_s)=L_z(t_c)+\int_{t_c}^{t_s}\dot L_z(t)\,\dd t,
 \qquad
 \Omega(t\ge t_s)=\frac{L_z(t_s)}{I(t)}.
 \label{eq:plateau}
\end{equation}
 The angular velocity can still change if the mass distribution inside the cylinder changes. 

\section{Conclusions}

We investigated a mechanism by which filaments in the local Universe acquire angular momentum in the presence of a nonzero cosmological constant. We focused on its effect on tidal trajectories and the finite-time angular momentum input into an open filament core. A constant, complete signed net input comprising both boundary flux and matter torque was shown to lead to linear accumulation of angular momentum.

In the finite-time example we considered, the cosmological constant delays the counter-rotating inflow, thereby reducing cancellation and increasing the retained rotational momentum. The effective angular velocity depends on the evolving moment of inertia and is distinct from the projected galaxy-velocity signal.

%\balance

\makeatletter
\let\bibcite\NAT@testdef
\makeatother

\end{document}